\documentclass[aps,twocolumn,nofootinbib,preprintnumbers]{revtex4}
\usepackage{graphicx, float}
\usepackage{hyperref}
\usepackage{amsmath}
\usepackage{amssymb}
\usepackage[usenames,dvipsnames]{xcolor}
\usepackage{physics}

\usepackage{etoolbox}
\apptocmd{\thebibliography}{\raggedright}{}{}

\def\beq{\begin{equation}}
\def\eeq#1{\label{#1}\end{equation}}
\def\eeqn{\end{equation}}
\def\beqa{\begin{eqnarray}}
\def\eeqa#1{\label{#1}\end{eqnarray}}
\def\eeqan{\end{eqnarray}}
\def\CR{\nonumber \\ }
\def\leqn#1{(\ref{#1})}

\def\Xp{X^\prime}

\def\stacksymbols #1#2#3#4{\def\theguybelow{#2}
    \def\vp{\lower#3pt}
    \def\sp{\baselineskip0pt\lineskip#4pt}
    \mathrel{\mathpalette\intermediary#1}}

\def\intermediary#1#2{\vp\vbox{\sp
     \everycr={}\tabskip0pt
     \halign{$\mathsurround0pt#1\hfil##\hfil$\crcr#2\crcr
              \theguybelow\crcr}}}

\def\gsim{\stacksymbols{>}{\sim}{2.5}{.2}}

\begin{document}

\title{Dark Matter as a Solution to Muonic Puzzles}

\author{Maxim Perelstein, Yik Chuen San}
\affiliation{Laboratory for Elementary Particle Physics, Cornell
  University, Ithaca, NY 14853, USA} 

\begin{abstract}
We propose a simple model in which dark matter particle exchanges mediate a new quantum force between muons and nucleons, resolving the proton charge radius puzzle. At the same time, the discrepancy between the measured anomalous magnetic moment of the muon and the Standard Model prediction can be accommodated, and thermal relic abundance of the dark matter candidate is consistent with observations. The dark matter particle mass is in the MeV range. We show that the model is consistent with a variety of experimental and observational constraints. 
\end{abstract}

\date{\today}

\maketitle

\section{Introduction}

Observational evidence for the existence of dark matter (DM) is overwhelming. While DM comprises most of the matter in today's universe, and contributes about 20\% of the total energy density, there is no known elementary particle that can account for it. Many candidate theories have been proposed, extending the Standard Model (SM) of particle physics to include one or more dark matter particles. In many theories, DM particles have potential experimental or observational signatures going beyond the purely gravitational effects that have been observed. However, no non-gravitational signature of DM has been conclusively established so far.  

In this paper, we propose that dark matter particles are directly responsible for explaining a long-standing puzzle in particle physics, the proton charge radius anomaly. The value of the proton charge radius measured using Lamb shift in muonic hydrogen~\cite{Pohl:2010zza,Antognini:1900ns} does not agree, at about $5\sigma$ level, with the value obtained from electron-proton scattering and electron hydrogen spectroscopy~\cite{Mohr:2008fa,Mohr:2015ccw}. A similar discrepancy was observed in muonic deuterium~\cite{muD_Pohl_expt}. We note that some recent data on $ep$ scattering~\cite{Xiong:2019umf} and H spectroscopy~\cite{Beyer:2017gug,Bezginov:2019mdi} are in better agreement with muonic hydrogen results, while others~\cite{Fleurbaey:2018fih} confirm the discrepancy. At present, the sources of disagreement among electronic experiments are not understood, and in this paper, we take the point of view that the anomaly is real and demands an explanation in terms of Beyond-the-SM (BSM) phenomena. (For previous works that proposed BSM explanations of the proton charge radius puzzle, see {\it e.g.}~\cite{TuckerSmith:2010ra,Batell:2011qq,Carlson:2012pc,Karshenboim:2014tka,Liu:2016qwd}.) 

The key ingredient of our proposal is the idea that loop diagrams involving light dark matter states can induce a new ``quantum" force between SM particles~\cite{Fichet:2017bng,Brax:2017xho}. Since the flavor structure of DM couplings to the SM is unconstrained, it is plausible that this new force may be felt by muons but not electrons. As will be shown below, such flavor-dependent quantum force can account for the proton charge radius puzzle, without conflict with any existing experimental constraints. At the same time, the dark matter particle responsible for this force can be a thermal relic consistent with the measured cosmological DM abundance, as well as with all known bounds on DM properties. The DM particle is predicted to have a mass of about 10 MeV, an interesting range from the point of view of direct detection experiments.         

Another prominent experimental anomaly involving the muon is the anomalous magnetic moment $a_\mu$, whose measured value differs from the SM prediction by about 3$\sigma$~\cite{Bennett:2006fi,Davier:2010nc,Hagiwara:2011af}. While non-perturbative SM contributions may account for some or all of this discrepancy, in this paper we take the point of view that it is real and requires a BSM explanation. It turns out that in our model, the effect of DM loops on $a_\mu$ is subdominant to the shift induced by the mediator particles which connect DM and SM sectors. We show that this shift can indeed account for the observed discrepancy, and demonstrate a set of parameters for which $a_\mu$ and proton charge puzzles are simultaneously solved, DM particle is a thermal relic with correct relic density, and all experimental and observational constraints are satisfied. It is remarkable that the simple model presented here can account for such a diverse set of data pointing to physics beyond the SM.  

\section{Model}

\begin{figure}[t!]
	\centering
	\includegraphics[scale = 0.33]{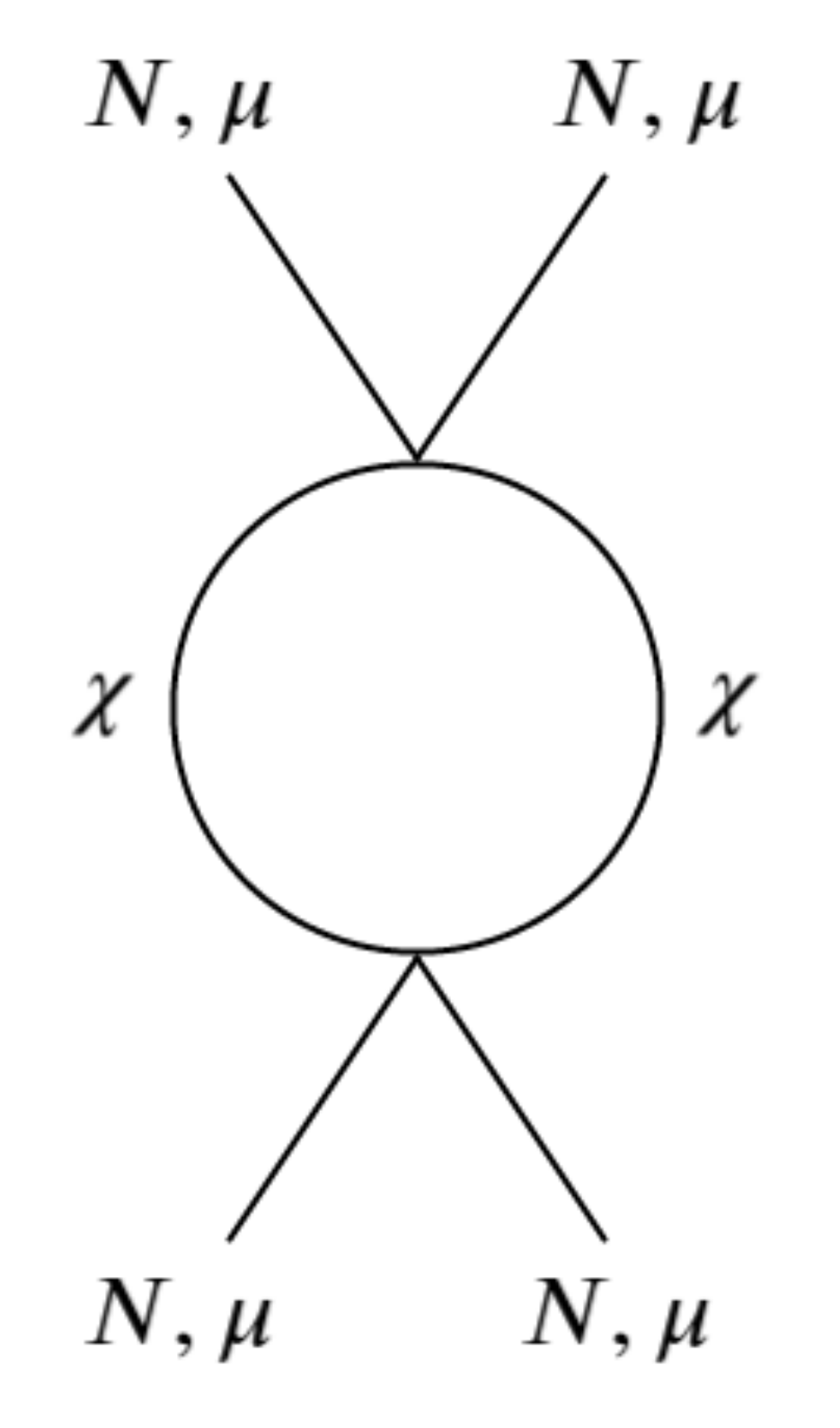}
	\caption{One loop diagram involving exchange of dark matter particle $\chi$ that induces a new force between muons and protons.}
	\label{fig:NewForce}
\end{figure}

We introduce a DM field $\chi$, a Dirac fermion with no SM gauge charges. DM is coupled to the SM via two real-scalar mediator fields, a leptophilic mediator $X$ and a leptophobic mediator $\Xp$. Both mediators are also SM gauge singlets. At energies below the QCD confinement scale, where all physics relevant for this study takes place, the interaction Lagrangian is given by
\beq
\mathcal{L}_{\rm int}
=
- g \bar{\mu}\mu X
-
(g'_p \bar{p} p + g'_n \bar{n} n) \Xp
-
y \bar{\chi} \chi X
-
y' \bar{\chi} \chi \Xp\,,
\eeq{Lint}
where $\mu$, $p$ and $n$ are muon, proton and neutron fields, respectively. Throughout this paper, we will study the regime in which both mediators are much heavier than the DM particle, $m_X, m_{X'} \gg m_\chi$, and can be integrated out, leading to an effective Lagrangian
\beq
\mathcal{L}_{\rm eff} =
         - \frac{yg}{m_X^2} \bar{\chi} \chi \bar{\mu}\mu
- \frac{y'g_p'}{m_{X'}^2} \bar{\chi} \chi \bar{p}{p} - \frac{y'g_n'}{m_{X'}^2} \bar{\chi} \chi \bar{n}{n} + \ldots
\eeq{Leff}
The quantum force between proton and muon, arising from the diagrams in Fig.~\ref{fig:NewForce}, provides a new contribution to the Lamb shift in muonic hydrogen, resolving the proton charge radius puzzle.  

A few comments are in order. For simplicity, we assumed that the mediator $X$ (and therefore the DM) couples to muon but not to the electron. While the flavor-dependent nature of this coupling is crucial to resolving the proton charge radius puzzle, a non-zero value of electron coupling (for example, a plausible scenario in which $g_\ell \propto m_\ell$) can be introduced without altering the basic picture. Further, $X^\prime$ is generically expected to couple to pions and other mesons. We do not include such couplings since they would play no role in our analysis. Finally, while the interactions in Eq.~\leqn{Lint} are sufficient to explain the proton charge radius and the $a_\mu$ anomaly, requiring that $\chi$ be a thermal relic with observed cosmological abundance necessitates an additional interaction involving neutrinos $\nu$:
\beq
\Delta {\cal L}_{\rm int} = -\lambda \bar{\nu} \nu X,
\eeq{DLint}
if neutrinos are Dirac, or its Majorana counterpart. This interaction can arise from the operator $X (HL)^2$ above the weak scale, and its strength is {\it a priori} unrelated to the coupling of $X$ to charged muons which arises from the operator $X (HL) \mu_R$.        

We also note that if $X$ and $\Xp$ were replaced with a single mediator, coupled to both muons and quarks, the dominant new physics effect in muonic hydrogen would come from a tree-level exchange of the mediator, rather than the DM-induced quantum force. To leading order, this model would in fact be identical to that already considered in~\cite{Liu:2016qwd}.   

\section{Results}

\begin{table}[t!]
	\centering
	\begin{tabular}{|c|c|c|c|c|c|c|c|c|}
		\hline
		 $m_\chi$& $m_X$ & $m_{X^\prime}$ & $g$ & $g_p^\prime$ & $g_n^\prime$ & $y$ & $y^\prime$ & $\lambda$ \\ \hline
		$9$~MeV & $25$~MeV & $45$~MeV & $4.5\cdot 10^{-4}$ & 0.01 & 0.01 & 0.5 & 0.5 & $10^{-4}$  \\
		\hline
	\end{tabular}
	\caption{Model parameters at the benchmark point.} 
	\label{tab:BP}
\end{table}

\begin{figure}[t!]
	\centering
	\includegraphics[scale=0.33]{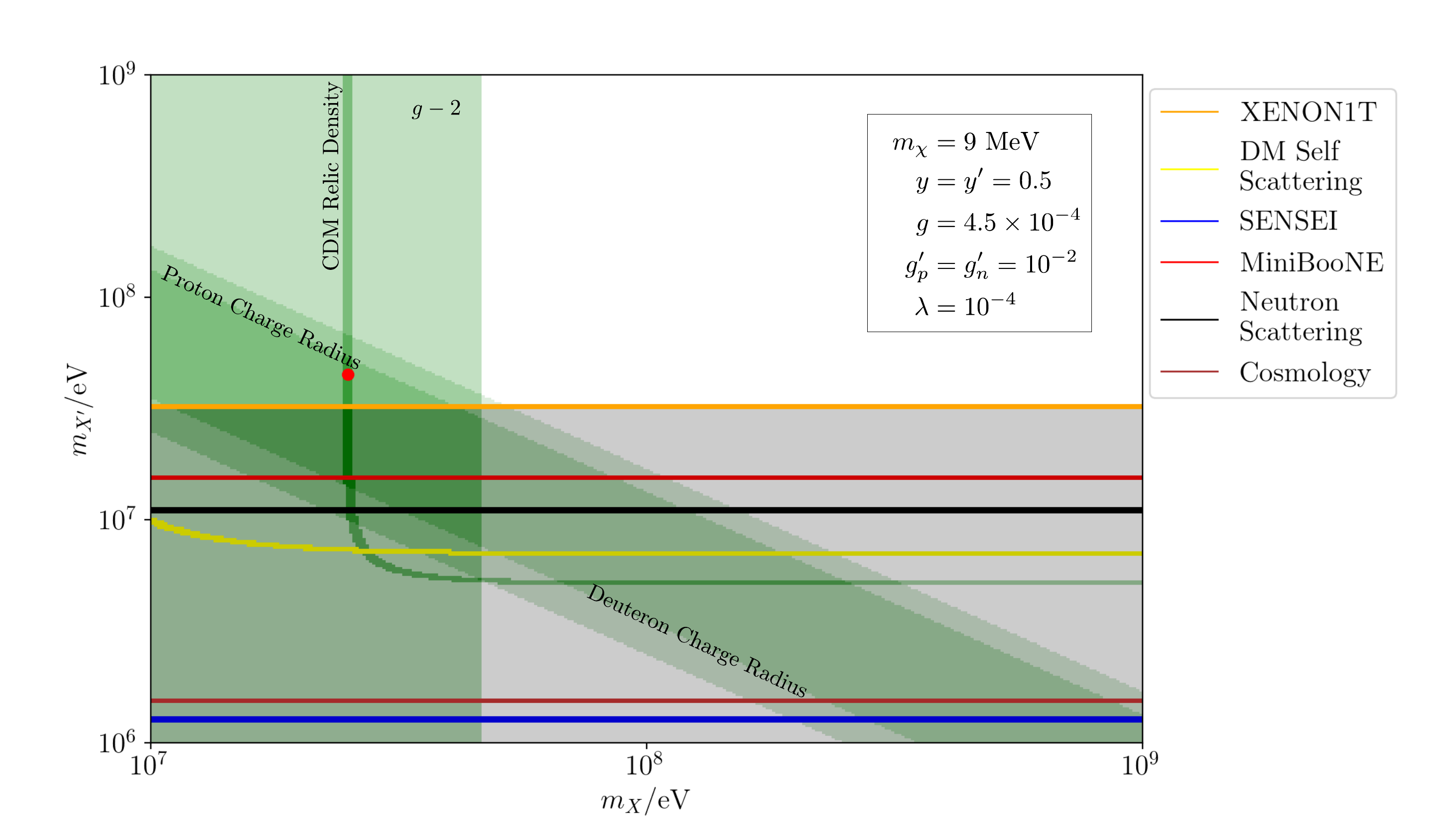}
	\caption{Fit to experimental data indicating non-SM physics (green) in the plane of mediator particle masses, $m_X$ and $m_{X^\prime}$, with the other parameters fixed to the values listed in Table~\ref{tab:BP}. Relevant experimental and observational constraints on DM and mediator particles are also shown; the shaded areas are ruled out.}
	\label{fig:mediator_masses}
\end{figure}

\begin{figure}[t]
	\centering
	\includegraphics[scale=0.33]{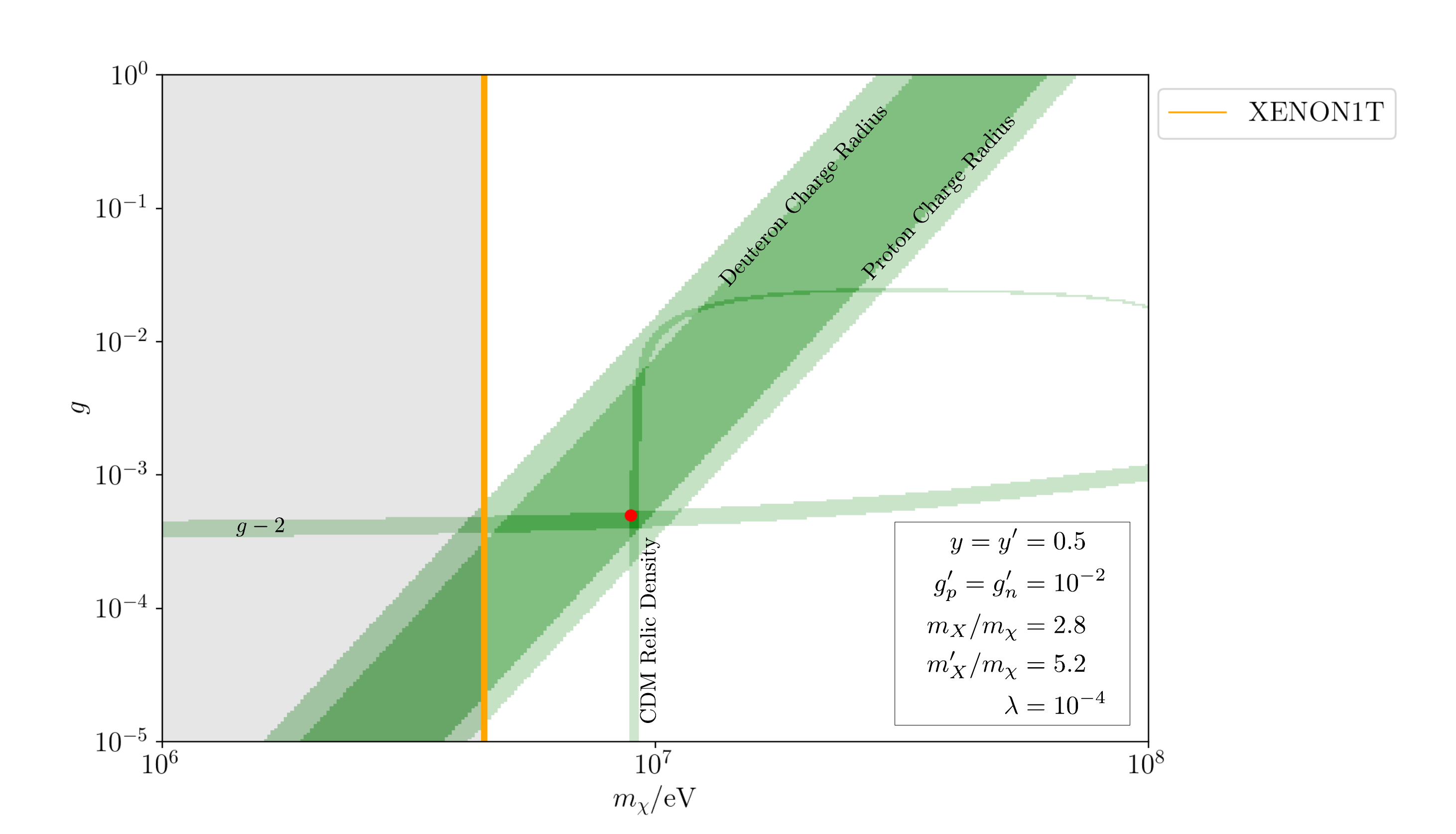}
	\caption{Fit to experimental data indicating non-SM physics in the plane of DM particle mass $m_\chi$ and the leptophilic mediator coupling to muons, $g$, with the other parameters fixed to the values listed in Table~\ref{tab:BP}.}
	\label{fig:varying_DM_mass1}
\end{figure}

\begin{figure}[t!]
	\centering
	\includegraphics[scale=0.32]{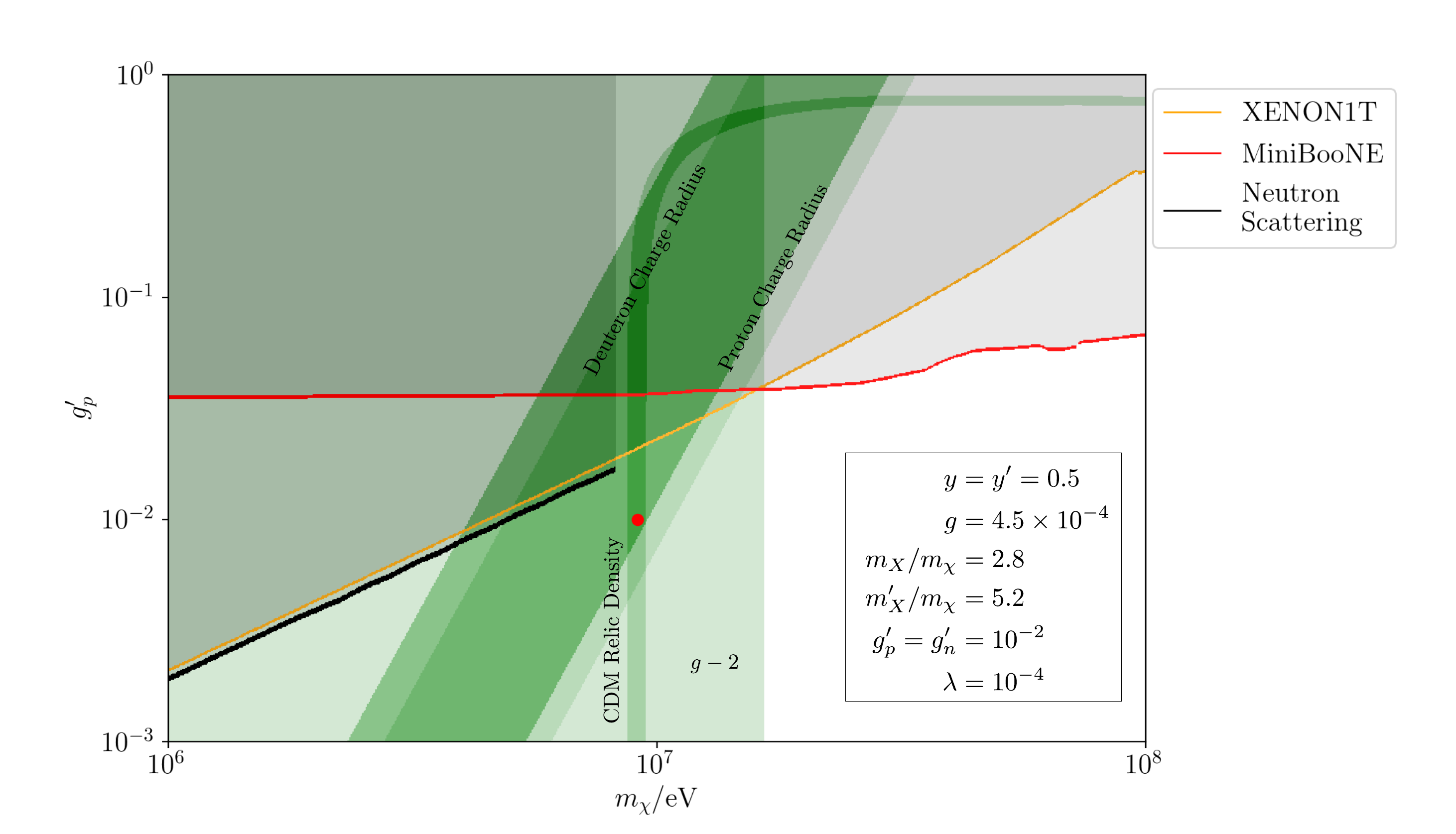}
	\caption{Fit to experimental data indicating non-SM physics in the plane of DM particle mass $m_\chi$ and the leptophobic mediator coupling to protons, $g_p^\prime$, with the other parameters fixed to the values listed in Table~\ref{tab:BP}. Relevant experimental and observational constraints on DM and mediator particles are also shown.}
	\label{fig:varying_DM_mass2}
\end{figure}

Using the above model, we performed a fit to relevant experimental data and observational constraints on DM properties. The results of the fit are summarized in Figs.~\ref{fig:mediator_masses},~\ref{fig:varying_DM_mass1} and~\ref{fig:varying_DM_mass2}. The model can explain the proton charge radius puzzle, the $a_\mu$ anomaly, and the observed DM relic density, while maintaining consistency with all known experimental and observational constraints. A sample benchmark point in the model parameter space which satisfies these requirements is shown in Table~\ref{tab:BP}. Some details of the analysis are presented below.  

{\bf Proton Charge Radius:} The fact that the proton has a finite size (with radius $r_p$) introduces shifts in energy levels of hydrogen-like atoms~\cite{Pohl:2010zza}. In particular, there would be a change in the energy difference between the $2S$ and $2P$ levels, {\it i.e.} Lamb shift $\Delta{E}_\text{Lamb}$. By measuring $\Delta{E}_\text{Lamb}$, one is able to deduce the value of $r_p$. In our model, the extra non-SM contribution to the Lamb shift in muonic hydrogen arises at one loop from the diagram in Fig.~\ref{fig:NewForce}. In the non-relativistic limit, this interaction can be captured by a potential between protons and muons~\cite{Brax:2017xho}:
	\beq
	V(r) = -\frac{3}{4\pi^3 r} \left(\frac{yg}{m_X^2}\right)
	\left(\frac{y^\prime g^\prime_p}{m_{X'}^2}\right)
	\frac{m_\chi^2}{r^2} K_2 (2m_\chi r),
	\eeq{eqn:DM_pot}
	where $K_2$ is the modified Bessel function of the second kind of order 2. As a result, there is a new contribution to $\Delta{E}_\text{Lamb}$ in muonic hydrogen, given by
	\beqa
	\Delta{E}_\text{L}
	&=&
	\bra{2S} V \ket{2S} - \bra{2P} V \ket{2P} \CR &=& 
	-\frac{3}{8\pi^3} \frac{yy'gg_p^\prime}{m_X^2 m_{X'}^2}\frac{m_\chi^2}{a^3}\,{\cal J} (x_0, a)\,.
	\eeqa{eqn:muH_Lamb}
Here
\beq	
	{\cal J}(x_0, a)\,=\,\int_{x_0}^\infty \dd{x}
	\frac{6\left(1-x\right)+x^2}{6x} e^{-x} K_2\qty(2m_\chi ax),
\eeq{Jdef}	
	$a$ is the Bohr radius of muonic hydrogen, and $x_0\approx \left(a\Lambda_\text{QCD}\right)^{-1}$ is the short-distance cutoff corresponding to the breakdown of the effective field theory description in Eq.~\leqn{Lint} at length scales below ${\cal O}(\Lambda_{\rm QCD})$. We do not include the additional contribution to the Lamb shift from length scales below $r_0$, which can only be calculated within a specific UV completion of Eq.~\leqn{Lint}. To estimate the associated theoretical uncertainty, we vary the value of the cutoff by a factor of two around the assumed central value $\left(a\Lambda_\text{QCD}\right)^{-1}$. This uncertainty is reflected in the width of the ``proton charge radius" band in Figs.~\ref{fig:mediator_masses}---\ref{fig:varying_DM_mass2}. Our fit assumes that this effect fully accounts for the ``extra" Lamb shift measured in the muonic hydrogen, compared to the baseline value inferred from electronic hydrogen and $ep$ scattering: $\Delta{E}_\text{L}=-0.307(56)$~meV. This is appropriate since in our model the electron does not couple to DM and hence is unaffected by the new force.    

\begin{figure}[t!]
	\centering
	\includegraphics[scale=0.33]{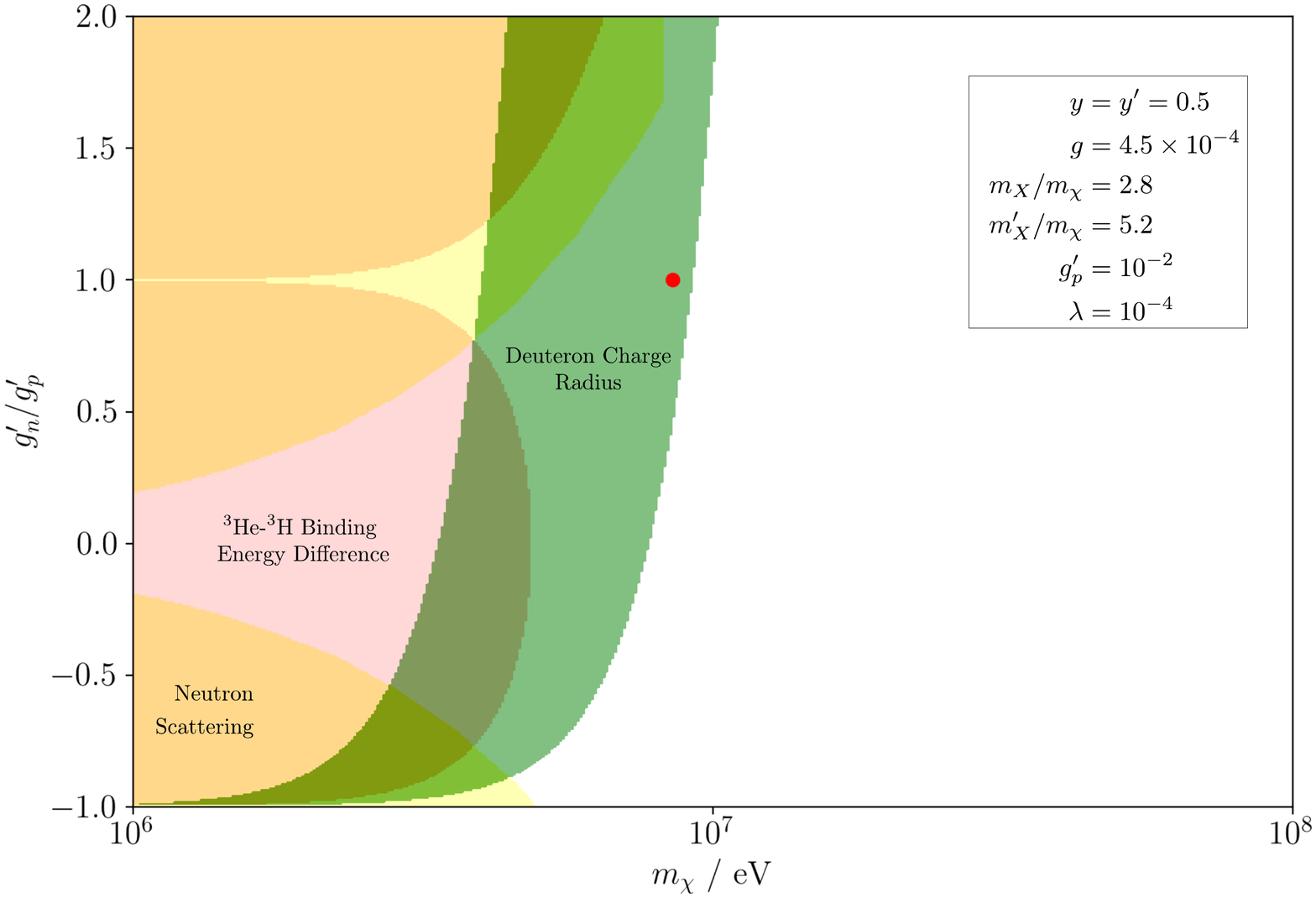}
	\caption{Fit to muonic deuterium data, as well as constraints from nuclear physics experiments, in the plan of the mediator coupling ratio $g^\prime_n/g^\prime_p$ and the DM particle mass $m_\chi$. The other parameters are fixed to the values listed in Table~\ref{tab:BP}.}
	\label{fig: g_prime_ratio}
\end{figure}

{\bf Muonic Deuterium:} After the proton charge radius puzzle was discovered, there has been interest in performing similar experiments with other muonic atoms, in particular muonic deuterium $\mu D$. It was reported that the deuteron charge radius $r_d$ extracted from $\mu D$ shows similar discrepancy from world-averaged CODATA 2014 value~\cite{muD_Pohl_expt},~\cite{Mohr:2015ccw}. When comparing against spectroscopic values of $r_d$ that involve deuterium only ({\it i.e.} independent of $r_p$), this discrepancy is reduced to $3.6 \sigma$~\cite{eD_radius}, but is still statistically significant. We therefore require that our model produces an extra contribution to Lamb shift in $\mu D$~\cite{muD_Pohl_expt}:
$\Delta{E}_\text{L}^{\mu D} = - 0.438(59) \text{ meV}$.
The Lamb shift in $\mu D$ is due to dark matter-mediated quantum force, and is given by
\begin{align} \label{eqn: muD_Lamb}
\Delta{E}^{\mu D}_\text{L}
=
-\frac{3}{8\pi^3} \frac{yg}{m_X^2} \frac{y'\qty(g'_p + g'_n)}{m_{X'}^2}
\frac{m_\chi^2}{a_{\mu D}^3}\,{\cal J}(x_{0D}, a_{\mu D})\,,
\end{align}
where $a_{\mu D}$ is the Bohr radius of the muonic deuterium system, $r_D$ is the deuteron radius, and $x_{0D}=r_D / a_{\mu D}$. The shift depends on both $g_n^\prime$ and $g_p^\prime$, while various combinations of these couplings are constrained by nuclear physics experiments (see below). As shown in Fig.~\ref{fig: g_prime_ratio}, isospin-preserving coupling $g_p^\prime=g_n^\prime$ is consistent with both the deuteron charge radius and nuclear physics constraints.  

	\begin{figure}[t!]
	\centering
	\includegraphics[scale = 0.35]{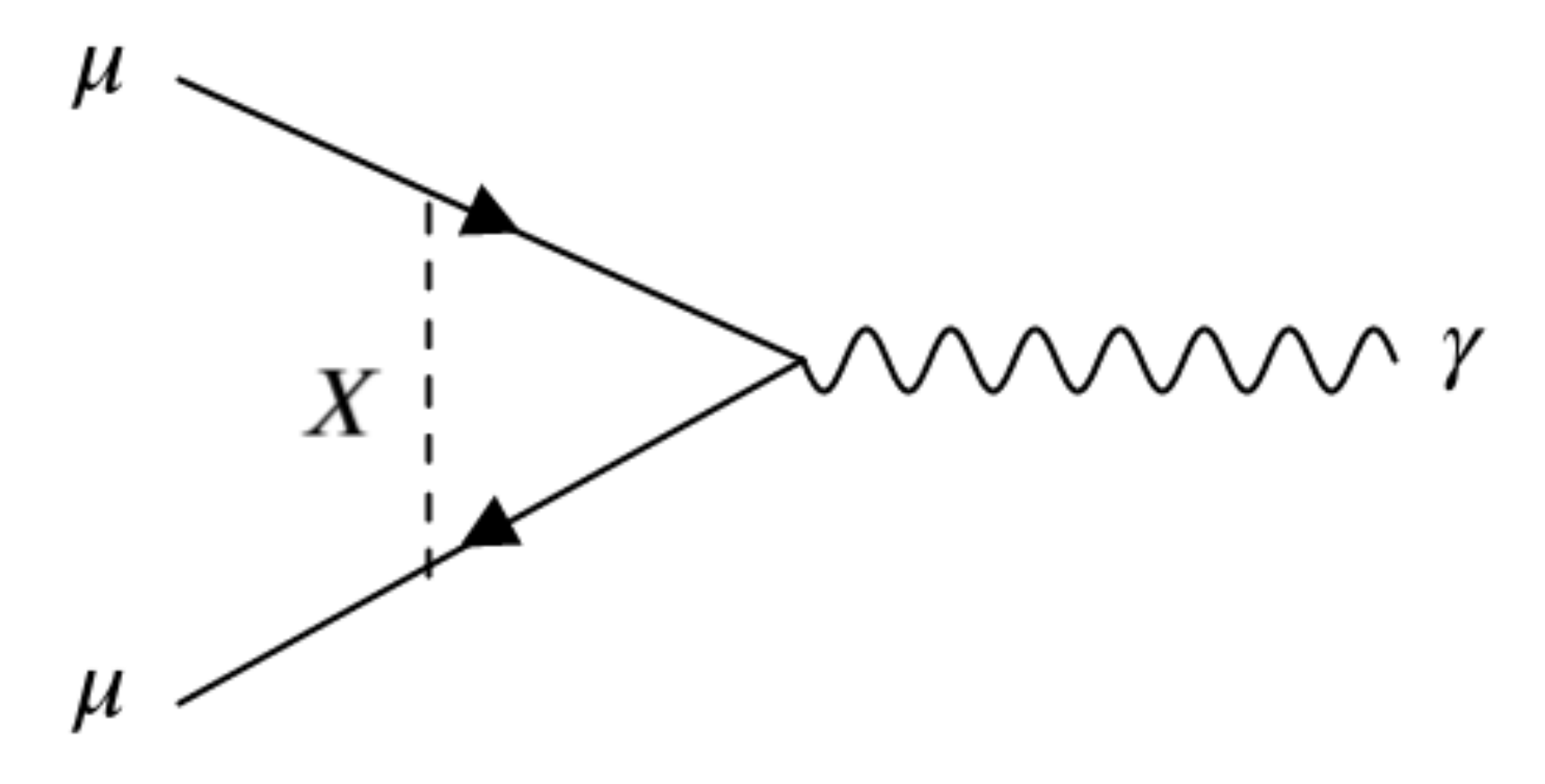}
	\caption{Contribution to muon anomalous magnetic moment due to the mediator particle $X$.}
	\label{fig: g-2_feyn}
\end{figure}

{\bf Muon Anomalous Magnetic Moment:} The leading new contribution to $a_\mu$ is given by the one-loop diagram shown in Fig.~\ref{fig: g-2_feyn}. Note that this contribution is independent of the dark matter candidate $\chi$ itself, which only enters at the two-loop level.
	The shift in $a_\mu$ is given by
	\beq 
	\Delta{a_\mu}
	=
	\frac{2g^2}{(4\pi)^2}
	\int_0^1 \dd{x} \;
	\frac{(1-x)^2(1+x)}{(1-x)^2 + x (m_X/m_\mu)^2}.
	\eeq{eqn:g-2}
	In our fit we assume that this effect fully accounts for the experimental discrepancy $\Delta{a_\mu} = 287(80) \times 10^{-11}$, within 2 standard deviations.

 \begin{figure}
	\centering
	\includegraphics[scale = 0.35]{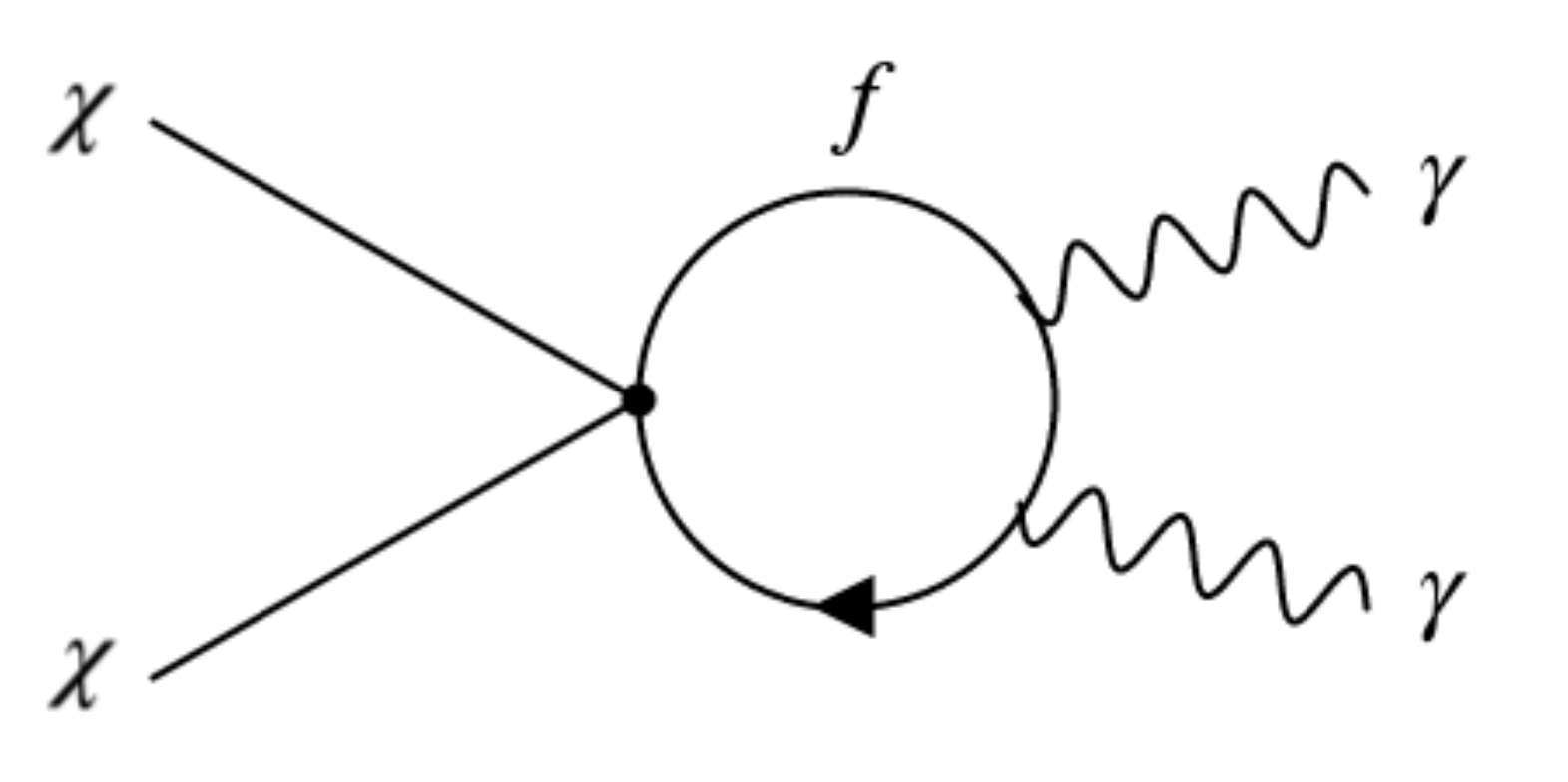}
	\caption{Dominant annihilation channel for $\chi$ at the time of freeze-out, if coupling to neutrinos are absent.}
	\label{fig:recli_decay}
\end{figure}

{\bf Relic Density:} We assume that $\chi$ is a thermal relic and that it accounts for all of the observed cosmological DM abundance, $\Omega h^2=0.120 \pm 0.001$~\cite{Aghanim:2018eyx}. With interactions in Eq.~\leqn{Lint} and $m_\chi \sim \order{\text{MeV}}$, the dominant DM annihilation channel at the time of freeze-out is $\chi\chi\to 2\gamma$, see Fig.~\ref{fig:recli_decay}. The leading ($p$-wave) contribution to the cross section in the non-relativistic limit is given by  
 \beq
\sigma_0
=
\frac{3e^2}{32\pi^3}\,\,
\abs{
	m_\mu I(\tau_\mu) \frac{yg}{m_X^2}
	+
	m_p I\qty(\tau_p) \frac{y'g_p'}{m_{X'}^2}
}^2.
\eeq{eqn:relic_cross}
Here the loop function $I(\tau_f)$ for a fermion $f$ is defined by
\begin{align} \label{eqn: loop_fun}
I(\tau_f)
=
\int_0^1 \dd{x} \int_0^{1-x} \dd{y} \;
\frac{1-4xy}{\tau_f - xy}
\end{align}
with $\tau_f = m_f^2/s$, where $\sqrt{s}$ is the center-of-mass energy of the scattering process. $\sigma_0$ can be used to compute the relic density of $\chi$ \cite{Kolb:1990vq} by solving the Boltzmann equation numerically.

We find that the $\chi\chi\to 2\gamma$ cross section is too small to provide the observed relic density for model paramaters required to fit the proton charge radius and $a_\mu$ anomalies. A simple solution is to consider an additional annihilation channel, $\chi\chi\to \nu\nu$, via the interaction in Eq.~\leqn{DLint}. The cross section is given by $\sigma_0
=\frac{3}{4\pi} \qty(\frac{y\lambda}{m_X^2})^2 m_\chi^2$. Since this final state arises at tree level, it naturally dominates over the $2\gamma$ channel.
With this addition, all three constraints can be satisfied simultaneously, see Figs.~\ref{fig:mediator_masses}---\ref{fig:varying_DM_mass2}.   	
	
In addition to fits to the data indicating deviations from the SM, a number of constraints from data and observations consistent with the SM have to be taken into account:

{\bf Dark Matter Self-Scattering:} Tree-level exchanges of mediator particles $X$ and $X^\prime$ induce DM short-range self-interactions of the form
    \beq 
{\cal L}_{\rm self}\,=\,-\qty(\frac{y^2}{m_X^2} + \frac{y'^2}{m_{X'}^2}) \qty(\bar{\chi}\chi)^2.
\eeq{eqn:dm_self}
DM self-scattering cross sections at low velocities are bounded by observations of halo shapes~\cite{alex2013snowmass2013,Zavala_2013}: $\sigma_T/m_\chi \lesssim 1 \text{ cm}^2/\text{g}$, where $\sigma_T$ is the momentum-transfer cross section defined by
\beq 
\sigma_T \equiv \int \dd{\Omega} \; \dv{\sigma_{\chi\chi\to\chi\chi}}{\Omega} \qty(1 - \cos{\theta}).
\eeq{mom_trans_cross}	
This constraint translates into an upper bound on the DM mass:
\beq 
m_\chi \lesssim 8\pi \qty(\frac{y^2}{m_X^2} + \frac{y'^2}{m_{X'}^2})^{-2} \times 4600~\qty(\text{GeV})^{-3}.
\eeq{eqn:dm_self_constraint}

{\bf Dark Matter Direct Detection:} Direct detection of dark matter in the MeV mass range has been the subject of much interest recently~\cite{Alexander:2016aln,Battaglieri:2017aum}. Most techniques rely on detection of DM scattering on electrons. In our model, this channel is not available, since by construction DM does not couple to electrons. However, scattering on a nucleon can occur, with cross section
\begin{align} \label{eqn: dm_nucleon_section}
\sigma\qty(\chi N \to \chi N)
=
\frac{1}{\pi} \qty(\frac{g_N'y'}{m_{X'}^2})^2
\qty(\frac{m_\chi m_N}{m_\chi + m_N})^2\,,
\end{align}
where $N=n$ or $p$. For our benchmark point, this cross section is about $6 \times 10^{-32}$~cm$^2$. This is about half an order of magnitude below the strongest current constraint from non-observation of signal due to energetic DM component generated through collisions with cosmic rays~\cite{Bringmann:2018cvk,Cappiello:2019qsw}, shown by the {\tt XENON-1T} curve in Fig.~\ref{fig:varying_DM_mass2}. MeV-scale dark matter can also be detected using the Migdal effect~\cite{Ibe:2017yqa,Dolan:2017xbu,Bell:2019egg,Baxter:2019pnz,Essig:2019xkx}. However, the recent results from {\tt SENSEI} collaboration~\cite{barak2020sensei} are not yet sensitive enough to constrain our model. 

{\bf Early-Universe Cosmology:} Measurements of Cosmic Microwave Backgound (CMB) place strong constraints on possible reionization due to DM annihilations~\cite{Slatyer:2015jla}. However in our model, $\chi$ annihilation proceeds in $p$-wave, and thus not subject to this constraint. CMB data together with the Lyman-alpha forest flux power spectrum from the Sloan Digital Sky Survey constrains the elastic scattering of DM on baryons~\cite{Boddy:2018kfv,Xu:2018efh}. This constraint is shown by the curve labeled {\tt "cosmology"} on Fig.~\ref{fig:mediator_masses}. In addition, a scenario where DM freeze-out occurs after neutrinos decouple from the rest of the SM plasma is constrained by the CMB bound on $\Delta N_{\rm eff}$, since in this case the neutrino temperature at recombination would be raised relative to $T_\gamma$ by the entropy transferred from the DM\footnote{We are grateful to Gordan Krnjaic for bringing this constraint to our attention.}. This argument imposes a lower bound $m_\chi\gsim$ a few MeV~\cite{Ho:2012ug,Boehm:2012gr}. A similar bound is imposed by the success of Big-Bang Nucleosynthesis (BBN)~\cite{Kolb:1986nf,Hufnagel:2017dgo,Nollett:2014lwa}.             

{\bf Dark Matter Mediator Searches:} In addition to direct searches for dark matter particles, there are many experiments looking for mediator particles produced at colliders or in a fixed-target setup~\cite{Alexander:2016aln,Battaglieri:2017aum}. While most analyses present the results in terms of bounds on dark photons, which couple to both leptons and quarks proportional to their electric charges, the interpretation of interest to us is in terms of leptophilic or leptophobic mediators. Leptophilic mediator searches rely on their production via their interaction with electrons, making them insensitive to our model. (An exception is the recently reported NA64 search for scalars produced through their coupling to photons~\cite{Banerjee:2020fue}. However, this search does not place relevant bounds on our model, since mediator couplings to photons are loop-suppressed.) Thus we only consider bounds from leptophobic mediator searches. With sub-MeV dark matter, the most stringent bound currently comes from the {\tt MiniBooNE} experiment~\cite{Battaglieri:2017aum,Aguilar-Arevalo:2018wea}, which places a bound on the parameter $Y$ related to dark matter annihilation cross section (see referenced papers for the precise definition). In our model, this parameter is given by
\begin{align} \label{eqn: miniboone Y}
Y = \qty(\frac{g_p'}{e})^2 \frac{{g_p'}^2}{4\pi} \qty(\frac{m_\chi}{m_{X'}})^2.
\end{align}
Note that the mediator $X^\prime$ in our model is a scalar, while the {\tt MiniBooNE} bounds were derived using a spin-1 mediator. An order-one correction to the bound may arise due to the differing kinematic acceptances and spin factors in the two cases, but we do not expect it to affect our conclusions.  	

{\bf Nuclear Interactions:} Leading non-SM contributions to nucleon-nucleon potential are given by
   \beqa 
V_{N_1 N_2}
&=&
-\frac{g'_{N_1} g'_{N_2}}{4\pi r} e^{-m_{X'} r} \CR & & \hskip-0.6cm
-\frac{3}{4\pi^3} \qty(\frac{y'}{m_{X'}^2})^2 
g'_{N_1} g'_{N_2} \frac{m_\chi^2}{r^3} K_2\qty(2m_\chi r).
\eeqa{eqn: 4N potential}
This extra potential can be probed at various nuclear physics experiments. It turns out that the second term in $V_{N_1 N_2}$, arising from the DM-loop exchange, is the most relevant for our analysis, since $1/r_{\rm exp} \ll m_\chi \ll m_{X^\prime}$, where $r_{\rm exp}$ is the length scale probed by the experiments. The relevant constraints are summarized in Fig.~\ref{fig: g_prime_ratio}. 

The binding energy difference between $^3$He and $^3$H has been well-established to be caused by Coulomb force and charge asymmetry of nuclear forces~\cite{Binding_diff_Coulomb},~\cite{Binding_diff_CSB},~\cite{Binding_diff_review}. In order not to spoil this agreement, the non-SM contribution is required to be less than 30 keV~\cite{Liu:2016qwd}. It is worth noting that this contribution is proportional to $\qty(g'^2_{p} -g'^2_{n})$, and vanishes in the isospin limit, see Fig.~\ref{fig: g_prime_ratio}. 

The charge-independence breaking (CIB) scattering length is defined as
\begin{align} \label{eqn: CIB}
\Delta{a} = \frac{1}{2} \qty(a_{nn} + a_{pp} - 2a_{np}),
\end{align}
where $a_{N_1 N_2}$ is the scattering length between two nucleons $N_1$ and $N_2$. Experimental and theoretical values for $\Delta{a}$ are known to be $5.64 \pm 0.60$ fm~\cite{CIB_scat_length_expt} and $5.6 \pm 0.5$ fm~\cite{CIB_scat_length_theory} respectively. Our model gives an extra contribution
\begin{align} \label{eqn: CIB_DM}
\delta{a}^\text{th}
= -\frac{3m_N}{\pi^2} \qty(\frac{y'}{m_{X'}})^2 m_\chi^2
\qty(g'_{p} - g'_{n})^2 \log\qty(\frac{m_\chi^2}{\Lambda^2}),
\end{align}
where $m_N$ is the nucleon mass and $\Lambda$ is the cut-off scale of the effective theory. We require $\delta{a}^\text{th} < 1.6$ fm to maintain agreement between experimental and theoretical values of $\Delta{a}$ at $2\sigma$ level. This constraint again vanishes in the isospin limit.

Scattering lengths between cold neutrons and nuclei can be measured by different methods, such as Bragg diffraction and the transmission method~\cite{neutron_scat_methods}. In our model, this scattering length is given by Eq.~\leqn{eqn: CIB_DM} by replacing $\qty(g'_{p} - g'_{n})^2$ with $g'^2_{n}$ or $g'_{n} g'_{p}$. By comparing the scattering lengths measured by different methods, bounds can be placed on contributions from non-contact operators. 
The detailed analysis can be found in~\cite{Nesvizhevsky:2007by}. Neutron scattering places the most restrictive nuclear-physics bound on the model in the isospin-symmetric limit~\cite{Brax:2017xho}.

\section{Conclusions}

In this paper we presented a simple model of MeV-scale Dirac fermion dark matter $\chi$ coupled to nucleons and muons, but not electrons. The quantum force due to $\chi$ loops is responsible for resolving the proton (and deutron) charge radius puzzles, while a scalar particle introduced to mediate DM-muon interactions can account for the discrepancy between the SM prediction for the anomalous magnetic moment of the muon and the measured value. If an additional interaction between $\chi$ and neutrinos is postulated, the former can be a thermal relic responsible for all of the observed DM abundance. We verified the existence of a region in the paramater space of the model where all of the above features are obtained simultaneously, while all known experimental and observational constraints on light DM and mediators are satisfied. 

New experimental data will soon be available that will test our model on various fronts. Numerous efforts, {\it e.g.} {\tt MUSE} experiment~\cite{Gilman:2017hdr}, are under way that will hopefully clarify the status of the proton charge radius puzzle. Anomalous magnetic moment of the muon measurement will be improved by the Fermilab Muon $g-2$ experiment~\cite{Grange:2015fou}. Experimental exploration of MeV-scale dark matter and dark sectors is an active and expanding area of research~\cite{Alexander:2016aln,Battaglieri:2017aum}. While our model has some amount of freedom in parameter choices, the available parameter space is finite and it is likely that this idea will be tested conclusively by the upcoming experiments in the near future.       

\acknowledgments{We are grateful to Gordan Krnjaic and Philip Tanedo for useful comments on the first version of this paper. This work is supported by the U.S. National Science Foundation grants PHY-1719877 and PHY-2014071.}

\bibliography{refs}
\bibliographystyle{apsper}

\end{document}